\documentclass[a4paper,11pt]{article}
\usepackage[T2A]{fontenc}
\usepackage[cp1251]{inputenc}
\usepackage{amsmath,amsthm,amssymb}
\usepackage{graphics}
\usepackage{graphicx}
\begin{document}
\title{Spin-Hall Detector}
\author{Yu.N. Chiang, M.O. Dzyuba\\
\emph{B. I. Verkin Institute for Low Temperature Physics and Engineering,}\\
\emph{National Academy of Sciences of Ukraine.}\\
\emph{Nauki Ave. 47, Kharkov 61103, Ukraine}\\
\emph{E-mail:chiang@ilt.kharkov.ua}}
\date{}
\maketitle

\begin{abstract}
An oscillographic study of the Hall voltage with an unpolarized alternating current through a platinum sample revealed choral features of the Hall effect, which clearly demonstrate the presence of the spin Hall effect in metals with a noticeable spin-orbit interaction. It was confirmed that, as in the case of direct current, the possibility of a spin-Hall effect is associated with the presence of an imbalance of the spins and charges at the edges of the samples, which is realized using their asymmetric geometry. In particular, it was found that such chiral features of the nonequilibrium spin-Hall effect, such as independence from the direction of the injection current and the direction of the constant magnetic field, in the case of alternating current, make it possible to obtain a double-frequency transverse voltage, which can be used as a platform for creating spintronics devices.

\end{abstract}
\section{Introduction}

  Since the introduction of the concept of an additional degree of electron freedom, the spin has been predicted and further a number of  characteristic properties of the electron wave function in its behavior, which follow from the relativistic Dirac equation, have been   described. So, in 1929, Mott first showed that one of them could be the chiral asymmetry of scattering of electrons with different spin directions in a central force field under conditions of a relativistic spin-orbit (SO) coupling [1]. After 40 years, Dyakonov and Perel, based on this idea, predicted for non-magnetic conductors the effect of curving electron trajectories with opposite spin orientations followed by their accumulation at opposite ends of the samples [2], which served as an impetus for active studies of the possibility of generating spin currents using Spin-Hall Effect (SHE). Since that time, many SHE experiments have been carried out in the framework of the concept of the Mott impurity mechanism of asymmetric spin scattering [3]. Later, under the conditions of SO interaction, which removes double spin degeneracy, the spin-dependent behavior of electrons in the absence of scattering was predicted due to the possibility of spin-dependent induction of the transverse electron velocity component in an external electric field  [4, 5, 6 ]. However, for a long time, the detection of these effects by electric methods in a “pure experimental setup” using an unpolarized injection current seemed impossible. To circumvent this difficulty, they resort to preliminary polarization of the current injected into the samples using ferromagnets, the ability of which to create spin polarization has been repeatedly confirmed, for example, by the manifestation of an anomalous spin-hall effect in them. The spin-polarized current obtained in a ferromagnet, which was then introduced into the material under study, induces a charge and spin imbalance in the sample (in particular, in SHE), which makes it possible to use electric measurement of spin currents [7, 8] as additive to the nonequilibrium state of charges and spins  in the system as a whole [9, 10], which leaves a certain ambiguity in the interpretation of the results.

  In [11], we proposed a method for creating nonequilibrium in spins and, correspondingly, in charges, which made it possible to study the momentum-spin dynamics of electrons in metals by electric methods, without resorting to improper methods of polarizing the current introduced into the sample. It consists in the use of samples with an asymmetric cross-sectional shape, the characteristic size of which  $ d\sim \sqrt {A_{yz}} $ (\emph {A} -sectional area) significantly exceeds the mean free path of carriers $ \ell_{c } $ and spin relaxation $ l_{sf} $.
\section{NSHE with alternating current}

  As is well known, the electrostatic potential difference in a metallic sample, including the transverse Hall voltage $U_{y}$ in the conventional Hall effect (CHE), is connected with the geometric dimensions of the sample, indicating that $U_{y} $ is determined by the total rather than the specific the number of charges in the sample (in contrast to the Hall constant). In other words, for an asymmetric cross section of the sample, $U_{y}$ depends on the formation of the distribution of all carriers along the cross section. If the corresponding gradient of the electrochemical potential $\mathcal {E}_{y}$ is established by the transverse current of all carriers under the action of the Lorentz force, as in CHE, then $U_{y}$ is determined simply by the mean section thickness $\bar{d}$ on axis \emph{z} passing through the center of gravity of the section ($U_{y}\sim\bar{d}^{-1}$). However, the dynamics of spins in \emph{a spin-orbit field}, as noted, can not lead to the same nonzero result for $\mathcal{E}_{y}$, since it generates equal, oppositely directed spin streams. Indeed, if the spin-orbit interaction, removing the energy degeneracy in the spin [9], leads to the appearance of the additive to the SO field $\mathcal{E}_{z}^{s}$, induced by the $\delta\mathbf{v}\neq 0$ drift additive to the carrier velocity due to the current $j_{c,x}$ along the \emph{x}, then  due to the multidirectional dynamics of the spins in the momentum space the deviation of electron spins for $ k_{y}> 0 $ up and $ k_{y} <0 $ down (\emph{k} - quasimomentum) will be antisymmetric and equal in magnitude, generating the equality of oppositely directed transverse currents relative to \emph{x} : $|+j_{y}^{s}|\equiv |-j_{y}^{s}|$.

   The process will end with the establishment of equilibrium between the strength of the spin-orbital field $F_{so}$ and the gradient of the spin chemical potential $\nabla\mu_{y}^{s}$, arising between the spins of the opposite direction ($F_{so}^{\pm} = e_{\mp} \nabla \mu_{y}^{s}$) and the accumulation of charges with opposite spins on opposite edges of the sample in equal amounts, regardless of the cross sectional geometry, if there are no spin relaxation processes. In this case, the appearance in the \emph{y} direction of the gradient of the spin chemical potential $\nabla\mu^{s}_{y}$ will not be accompanied by a charge imbalance and the gradient of the electrochemical potential in this direction remains zero. This means that the condition $\ell_{sf}\gg\sqrt {A_{yz}}$ makes it impossible to study SHE by electrical methods.

    However, if the dimensions of the sample in the cross section significantly exceed the mean free path ($ \sqrt {A_{yz}}\gg\ell_{c} $), so that the inequalities $\ell_{c}\ll\ell_{sf }\ll\sqrt {A_{yz}} $, then the "spin-flip" region is inevitable,where the dynamics of spins is stochastic, and the spin currents of oriented spins will appear only where the spins are coherent. In Fig. 1, the regions of coherent spin flows in the field $ \mathcal {E}_{y}^{z} $  near the opposite \emph{a} and \emph{b} edges of the sample with thicknesses $ t_{a} $ and $ t_{b} $ in the \emph {z} direction differ ($ t_{a}> t_{b} $) and have an unequal number of carriers (spins) $ N_{a} \sim A_{a} $ and $ N_{ b} \sim A_{b} $. Arrows crossed out by crosses indicate spin flows that do not reach the edges of the sample. As a result, the total charge current in the \emph {y} direction due to spin dynamics under conditions of SO interaction will  be as follows:

    \begin{equation}\label{1}
 \overline{I}_{y}^{s}=I_{a}^{s}-I_{b}^{s} = I_{a}^{s}(1-\frac{I_{b}^{s}}{I_{a}^{s}}) = I_{a}^{s}(1-\frac{N_{b}}{N_{a}}) = \sigma_{xy}^{s}\mathcal{E}_{y}^{z}(1-\frac{A_{b}}{A_{a}})
\end{equation}
($\sigma_{xy}^{s}$ is spin conductivity).
\begin{figure}[htb]
\begin{center}д
  \includegraphics[width=10cm]{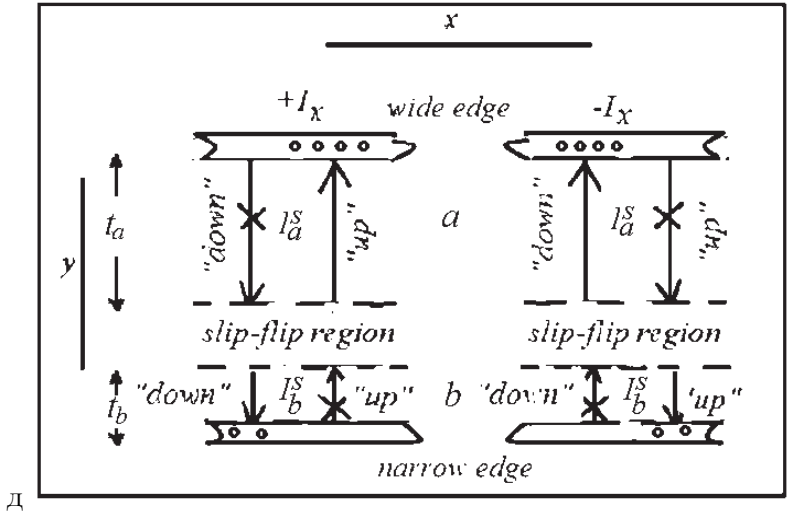}\\
  \caption{The diagram of spin-currents in an asymmetric sample with a finite length of spin relaxation. \emph{a} and \emph{b} areas  refer to $ N_{a}$ and $ N_{b}$ numbers of carriers, respectively. }\label{1}
  \end{center}
\end{figure}

Thus, in an asymmetric sample, spin imbalance is accompanied by charge imbalance, which allows one to study the features of the spin-hall effect by electric methods without the aid of ferromagnets. The distinctive features of such a nonequilibrium SHE at constant current, studied in detail in [12], should be the independence of the direction of nonequilibrium gradients of the spin and charge chemical potentials from the direction of the current, as follows from the diagrams in Fig. 1, and in a constant magnetic field and from the direction of the field: Unlike CHE, which is a vector quantity, the interaction of oppositely oriented spins '\emph {up}' and '\emph {down}' with a magnetic field in opposite directions should be symmetrical, so that the spin conductivity should be scalar.

The indicated NSHE properties make it easy to separate the Hall components of the voltage transverse to the current and field directions (see [12]) — spin $ U_{\rm SHE}$ and Lorentz $\mathbf {U}_{\rm CHE}$:

\begin{equation}\label{2}
  \mathbf{U}_{y}=U_{\rm SHE} + \mathbf{U}_{\rm CHE} = W\overline{j}_{y}^ {s}/\sigma^{s}(B) + U_{\rm CHE}[\mathbf{x}\times\mathbf{z}] \rightarrow U_{y} = U_{\rm SHE} \pm  U_{\rm CHE} ,
    \end{equation}
 where $\mathbf{x}~\mbox{и}~\mathbf{z}$  are orts of axes \emph{x} and \emph{z} ; $\overline{j}_{y}^ {s}$ and $\sigma^{s}$ are spin current density and spin conductivity, respectively;  the \emph {W} is the average distance between the sample edges  not  equal thickness along \emph {y}, and the signs of the second component correspond to the orientations of the magnetic field of the opposite direction.

   In this report, we show how these features of NSHE using alternating current can be demonstrated visually and used as an informative platform for elements of spinelectronics, such as spin detectors, for example.
     Indeed, since, according to the diagram in Fig. 1, the sign of the spin-charge unbalance does not depend on the direction of the current, then the possibility of an oscillographic visualization of the effect should obviously follow from this fact.
     With alternating current, for example, sinusoidal form $ J_{x} = J_{x0} \sin \omega t $ (hereinafter the symbol 0 will denote amplitude), expression (1) will represent the rule for determining the total amplitude of the transverse amplitude-modulated signal $ F_{am} $ consisting of a 'carrier', $ U _{\rm SHE} (t, B) = U_{\rm SHE (0)} (B) f _{\rm SHE} (t) $, and ' modulating ', $ U_{\rm CHE} (t, B) = U _ {\rm CHE (0)} (B) \sin \omega t $, components:
      \begin{eqnarray}\label{3}
      F_{am}^{y} = F_{am (0)}(B)f_{\rm SHE}(t) & = & [U_{\rm SHE(0)}(B) + U_{\rm CHE}(t, B)]f_{\rm SHE}(t)  =
      \nonumber\\
      && = U_{\rm SHE}(t, B) [1 \pm \alpha_{m}(B)\sin\omega t ],
            \end{eqnarray}
          where  $\alpha_{m}(B) = \frac{U_{\rm CHE (0)}(B)}{U_{\rm SHE (0)}(B)}$  is modulation factor (depth).
     The expression shows that if NSHE ($ U_{\rm SHE} $) has the property not to depend on the directions of the current and the magnetic field, then the first component in this expression should be an even function of the current ($ f_{\rm SHE} (t) \sim f (J_{x}^{2})) $ and the \emph{B} sign, and the second component is odd both in current and \emph{B}, being a product of functions of different parity, and therefore, both components should visually differ in the form and duration of periods (2 times). This allows direct oscillographic visualization of the effect on alternating current for certain values of the modulation depth $ \alpha_{m} (B) $.

     \begin{figure}[ht]
\begin{center}
  \includegraphics[width=10cm]{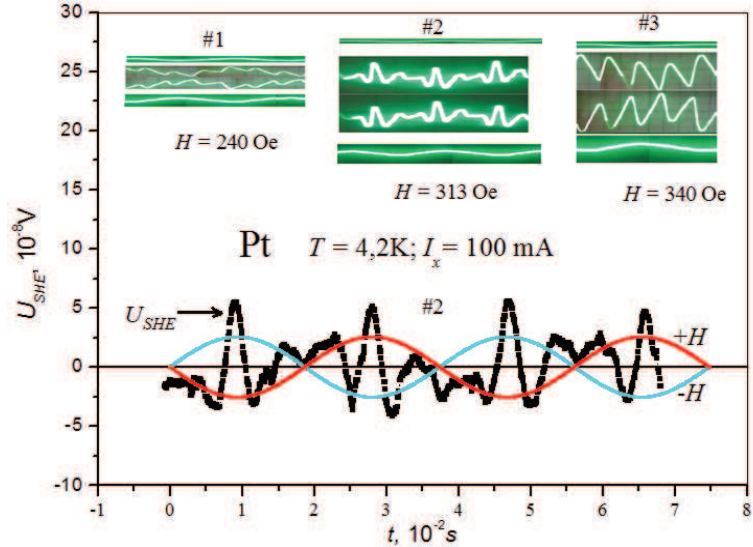}\\
  \vspace{-0,5cm}
  \caption{Nonequilibrium Spin-Hall Effect at sinusoidal injection current. The inserts are oscillogram series with different modulation depths of the spin-hall effect ($U_{\rm SHE}$)by the normal Hall effect ($U_{\rm CHE}$). Below are signals of series 2: $U_{\rm SHE}$ is signal SHE double frequency ($2\omega$); +H, -H are $U_{\rm CHE}$ modulating  signals with a frequency equal to the current frequency ($\omega$).}\label{2}
  \end{center}
\end{figure}
\section{Visual implementation of the NSHE}
     
      To implement the visualization procedure, we chose a heavy metal (Pt), as a metal with the expected strong spin-orbit interaction and, as a result, a sufficiently large spin-hall effect for the reliability of its resolution, since the alternating current requires the use of broadband non-selective recording mode.
           
           Samples were prepared in the form of pieces of rolled foil sized $ 6.5 \times 2.5 $ mm $^{2} $ with edge thicknesses differing by $ \approx 0.1 $ mm, which was not less than 300 electron mean free paths $ \ell_{c} (\approx 0.3 \mu $ at 4.2 K). Signal resolution \emph{ac} was $ \sim 10^{-9} $ V.

        The insets in Fig. 2 show the oscillograms of $ U_{y} $ for different values of the  AC sinusoidal modulation coefficient with an injection current amplitude of the $ I_{x0} = 100 $ mA and a frequency of 22 Hz. The three most characteristic waveform series are shown: with $ \alpha_{m} (B) \gg $ 1 with small values of $ U_{\rm SHE (0)} (B) $ and $ U_{\rm CHE (0)} (B) $ (series $\#1$) and large values of $ U_{\rm CHE (dc)} (B)$ 1(series $\#3$), and for $ \alpha_{m} (B) $ <1 (series $\#2$). In each series, oscillograms are presented for two opposite directions of the magnetic field (middle photos), which were realized by switching the direction of the current by changing its phase by $ \pi $. At the top of each series are initial waveforms at \emph {H} = 0, associated with a slight non-orthogonality of the Hall contacts.

        Separately shown is the $ U_{\rm SHE} $ curve from series 2, averaged over the two waveforms shown in the $ \# $ 2 insert for $ \pm B $ after excluding $ U_{\rm CHE} (t, B) $ according to the rules of expression 1. It is seen that the phase $ U_{\rm SHE} $ does not depend on the phase of the modulating signal, as expected for the spin-Hall effect.

        The oscillograms obtained with the same instrument resolution and the same scanning frequency. The bottom waveforms on the inserts (modulating envelopes) demonstrate the dependence of the modulation amplitude on the magnetic field. For convenience of analysis, the scanning frequency was chosen such that on all photos the curves           were represented within at least a period. To eliminate the influence of the field on the electron dynamics, small fields were used that satisfy the condition $ \omega_{c} \tau \ll 1 $, where $ \tau = \ell_{c} /v_{\rm F} $ is the momentum relaxation time, $ v_{\rm F} $ is a Fermi-velocity.
             Starting with some values of the magnetic field, peaks appeared on the oscillograms, whose order and sign, the same for opposite field directions, corresponded to the features of the even function relative to the sign of the current and magnetic field.
      
       In addition, when changing the direction of the magnetic field, the phase of the modulating signal changed as $ \pi $ as an alternating (odd in current and magnetic field) contribution $ U_{\rm CHE}$.
            Thus, the observed features of the oscillograms completely corresponded to the manifestation of NSHE: the spin-hall effect “straightens” (!) the current, as the series of oscillograms $ \# 2 $ particularly clearly demonstrates.
                Obviously, the possibility of such a visual visualization of the effect is limited to the range of the modulation factor $ \alpha_{m} (B) <$ 1. In the case of $ \alpha_{m} (B) \gg $ 1, the type of even-current effects is detected after digital processing data by formulas 1 by averaging the data for opposite directions of the magnetic field. The resulting dependence of the amplitude of the AC response of the double frequency NSHE in the entire measured range of magnetic fields has the form shown in Fig. 3.

        \begin{figure}[htb]
\begin{center}
  \includegraphics[width=10cm]{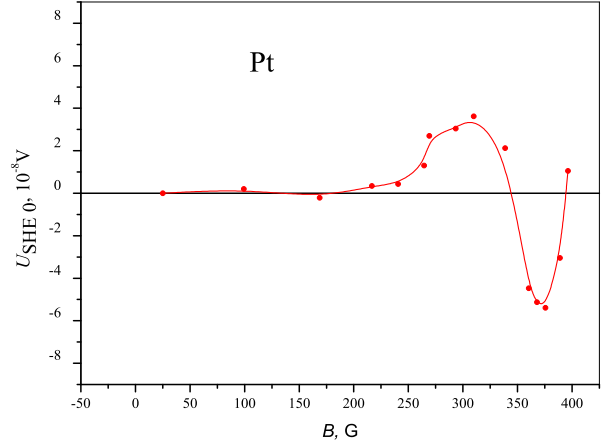}\\
  \caption{The resulting dependence of the amplitude of the AC response of the double frequency NSHE.}\label{3}
  \end{center}
\end{figure}
                It can be seen that sign of the amplitude $ U_{\rm SHE (0)} (B) $ of this response changes  to the opposite at some point of the crossover, which corresponds to the change of the AC phase of the signal by $ \pi $. As in the case of Al, this is apparently due to a change in the hole carrier sign [12]. According to the measured data, the spin Hall angle for platinum is $ \tanh\theta_{SH} \approx 10^{- 2} $, which is almost an order of magnitude smaller than the usual Hall angle.
\section{Conclusion}
      
        In conclusion, with the help of alternating current, we visualized a nonequilibrium spin-hall effect in asymmetric platinum plates in the helium temperature range. The observed oscillograms completely confirmed the properties of the spin-hall effect, which we previously discovered when using unpolarized direct current injection into samples of metals such as Al, W, Pt.The chiral properties of the effect on alternating current make it possible to use them as an informative platform for simple devices of spinelectronics that do not require high technologies for the production of samples.


\begin{thebibliography}{99}
  \bibitem{1}Mott N.F.\emph{Proc.Roy.Soc.} \textbf{A124}, 425 (1929).
  \bibitem{2}M.I. Dyakonov, V.I. Perel. \emph{Pisma v JETP} \textbf{13}, 657 (1971)[\emph{Phys. Lett.}, \textbf{A 35}, 459 (1971)].
  \bibitem{3}Jairo Sinova, Sergio O. Valenzuela, J. Wunderlich, C.?H. Back, and T. Jungwirth. \emph{Rev. Mod. Phys.} \textbf{87}, 1213 (2015).
   \bibitem{4}Bychkov Yu.A. and Rashba E.I. \emph{Pisma v Zh. Eksp. Teor. Fiz.} \textbf{39}, 66 (1984).
  \bibitem{5}S. Murakami, N. Nagaosa, and S.-C. Zhang. \emph{Science} \textbf{301}, 1348 (2003)
  \bibitem{6}Sinova J., D. Culcer, Q. Niu, N. Sinitsyn, T. Jungwirth, and A. H. MacDonald. \emph{Phys. Rev. Lett.} \textbf{92}(12), 1(2004).
   \bibitem{7}Valenzuela S. O., and M. Tinkham. \emph{Nature} \textbf{442} (7099), 176 (2006).
  \bibitem{8}F.J. Jedema, H. B. Heersche, A. T. Filip, J. J. A. Baselmans, and B. J. van Wees, \emph{Nature} (London) \textbf{416}, 713 (2002).
  \bibitem{9}R.V. Shchelushkin and Arne Brataas. \emph{Phys. Rev. B} \textbf{72}, 073110 (2005).
  \bibitem{10}Zhang, S. \emph{Phys. Rev. Lett.} \textbf{85}, 393 (2000).
  \bibitem{11}Yu.N. Chiang and M.O. Dzyuba.\emph{EPL}, \textbf{120}, 17001 (2017).
 \bibitem{12}Yu.N. Chiang and M.O. Dzyuba. \emph{Physica B: Condensed Matter} \textbf{558} 44 (2019).
   \end{thebibliography}
\end{document}